# Maintaining The Humanity of Our Models

**Umang Bhatt**

Carnegie Mellon University
5000 Forbes Ave
Pittsburgh, Pennsylvania 15213
umang@cmu.edu

### Abstract

Artificial intelligence (AI) and machine learning (ML) has been a major research interest in computer science for the better part of the last few decades. However, all too recently, both AI and ML have rapidly grown to be media frenzies, pressuring companies and researchers to claim they use these technologies. As ML continues to percolate into the layman's life, we, as computer scientists and machine learning researchers, are responsible for ensuring we clearly convey the extent of our work and the humanity of our models. Regularizing ML for mass adoption requires a rigorous standard for model interpretability, a deep consideration for human bias in data, and a transparent understanding of a model's societal effects.

## Introduction

Mainstream media, any non-academic or non-research outlet, fawns over the tandem of machine learning (ML) and artificial intelligence (AI). The media is correct in claiming that, while ML is outperforming humans at clerical and pattern-driven work, the next wave of AI will revolutionize medicine, law, finance, and transportation by processing data more efficiently than humans (Grace and Salvatier 2017). It is not wrong to be proud of and eager about the advances made in these fields annually, but it is erroneous to overstate these technologies' capabilities in the immediate future, which we define as ~1-2 years. AI growth is slowly yet drastically automating aspects of the monotony in our lives (Schwab 2016). To clarify, in this context, we define AI as ML used to understand patterns in data and as the intelligent agents used to augment daily life.

As AI enters the limelight and displaces all regardless of the color of their collar, we, as researchers and practitioners of the field, must poise our models to be interpretable, unbiased, impactful, and thus humane (Kaplan 2015). We define humanity and humane to be the ethereal and emotional impact of models on humans. In order to build a system that values humanity, we must consider the following questions: (1) How can we make our work interpretable by the end user? (2) How can we ensure our algorithms are not learning patterns from yesteryear's potentially biased data? (3) How can we evaluate the societal effects of our predictions?

These three questions provide the foundation needed to succeed in maintaining the humanity of the models we create. To scale to the masses, our AI systems need be interpretable to a layman with no prior computer science knowledge. Laymen should be able to understand the sequence of steps and data points used (and their weights) to achieve the final result. Our systems must draw from unbiased training data. This is a largely over looked portion of current ML work: most researchers claim themselves to be data-agonistic, even though it is important to care about the features, source, and context of the datasets we train on (O'Neil 2016). Finally, our systems must be aware of the user impact of each prediction we make and pattern we find. Having a pointed, narrow goal with low impact is the current rule of thumb to ensure little disruption in other parts of a user's life (Armstrong and Levinstein 2017). To that end, we dive into the need for all three pillars, as our field's research continues to evolve at lightning speeds.

## Model Interpretability

Imagine a patient visiting a doctor's office in 2030. Picture that patient walking into an empty room with sensors and a large screen telling the patient what to do. Once the minimum readings have been made (non-invasively and implicitly), the patient can see a diagnosis generated by our black box, say in this case "Diabetes". Predicting and delivering a life-changing diagnosis in such an insensitive manner can stifle the adoption of our AI systems, since our system lacks humanity in diagnosis. Therefore, we must prioritize exposing the inner workings of our system to promote in-terpretability - the explanation behind our prediction – thus bringing us closer to more personable, humane models.

**Current State**

ML today begets a robust strength in prediction power in decision-making processes (at least in the supervised case, which we will assume from here). However, due to a mismatch between prediction objectives (i.e. test set performance) and the real world costs of deployment, there is an unfulfilled demand for interpretability (Lipton 2017). Laymen render models lacking interpretability effectively useless since it near impossible to follow knowledge extraction of those models. Though there exists no concrete definition of interpretability, it broadly refers to explaining a model in humanly understandable terms: many desiderata for modern ML systems, like robustness, fairness, and trust, are also commonly grouped with interpretability (Doshi-Velez and Kim 2017).

There is an exigency for a rigorous standardization for interpretability, since the European Union will prevent automated individual decision-making by 2018 (Goodman and Flaxman 2016). As of now, dimensionality reduction techniques like backward feature selection on a single layer perceptron or feature extraction via principle component analysis suffices to make a model interpretable in simple cases (Vellido 2012). Sparse linear classifiers and discretization methods (decision trees, rule sets, etc.) are well-known interpretable models (Kim 2015). However, much interest now lies in the nonlinear, high dimensional models and related deep learning techniques. In these situations, we want to provide laymen with explanations via known interpretable models

More recent techniques have actually implicitly prioritized interpretability, albeit void of a standardization. Researchers working on neural modulation for semantic search in visual content and those working on joint model training are inherently making ML models more interpretable by exposing the modular objects within the whole image and the sequence of states, respectively.

**Case Study: Medicine**

Returning to our 2030 scenario, the patient demands an explanation of how our complex model, like Doctor AI, came to its diagnosis. We know with a high degree of confidence that our prediction is correct; however, now we must expose the sequence of decisions that led to our conclusion. One option would be jointly training a recurrent neural network, a long short-term memory (LSTM) per se, with a hidden Markov model (HMM) to expose the HMM state sequences to the end user (Krakovna and Doshi-Velez 2016). This technique allows us to leverage both the predictive power of an LSTM and the explicit states of an HMM. We even unlock the power of transfer learning as an LSTM model trained on a sufficiently large electronic health record can be transferred to any hospital (Choi, Bahadori 2016). However, a major shortcoming of this approach is that a domain expert must be leveraged to name the states of the HMM: it is near impossible for a computer scientist to attempt to name a given state sequence of symptoms and vital signs as potentially contributing to a given diagnosis. In some simpler planning tasks, expert knowledge is taken into account in the prior distribution over the area of interest, but this does not generalize well to all situations (Kim 2015). Nonetheless, coupling combined model training with evaluating test set performance for the top-k ICD-9 codes[1] can produce accurate and interpretable results (Lipton and Kale 2015, Nigam 2016). Another such technique for making these predictively powerful LSTMs more explainable is employing input gradients to generalize decision logic, which is irrespective of the dataset (Ross, Hughes, Doshi-Velez 2017). These techniques are all means towards the end of making our ML models more interpretable and thus more humane.

## Human Bias in Data

The source and features of data used as a basis for our models are essential to understanding the inherent human bias in our predictions. When productionalizing a model, we must divulge the exact source and features of the data used to train that model. Data, contrary to layman's thoughts, does age and grow stale. Imagine if we used data from the Jim Crow days to predict in which zip codes are people most likely to go to jail again (O'Neil 2016). Overtime, the data from yesteryear is no longer relevant. So, can we not just create a threshold or add a layer of logistic weight to our data by recency? Well, a recency bias is just as unproductive (Abah 2016). Acknowledging the existence of and taking steps to correct this potentially unfair data yields more humane models, since an unbiased model fed biased data still gives a biased result.

**Current State**

When assessing the quality/recency of and reducing the bias of a dataset, two techniques are common. One technique is debiasing, which manually severs the learned relationship between two entities. In example, gender bias in natural language generation from processing/training on text corpuses is all too common. A gender bias-free dataset of images can be created when we place constraints on certain relationships between entities within the images (Zhao, Wang 2017). In a text generation algorithm, gender bias can be mitigated by identifying known gender biased words, working in a gender neutral subspace, and understanding the distance of a gender neutral world towards the preidentified gender subspaces (Bolukbasi, Chang 2016). Another technique is simply omission of the stale or biased data from training; it is trivial to state, but such a decision is lossy and certain patterns in the data will be missed.

---

[1] We pick the top k most frequent ICD-9 (alphanumeric codes for patient diagnosis) and classify the accuracy of our model on those codes.

It is crucial to note that in both scenarios, it is we, the researchers, imposing our own bias and morality on a given problem space. For example, if we think (or even empirically show) that zip code of residence is a high predictor of where crime occurs, we are then faced with a moral struggle of whether or not to patrol more in those zip codes, disadvantaging the portion of non-criminals in a zip code deemed crime prone. The legality of our model matters considerably as an ounce of human bias can violate the law (Samek 2017). To that end, we want to remove human bias disparities with as little impact on accuracy as possible (Johndrow and Lum 2017).

**Case Study: Recidivism**

Recidivism prediction, propensity of a person to return to jail once released, is bursting with human bias. Though models like PredPol exist, there is no formal feedback loop for all involved parties, and thus, we lack randomness in the data we use (Ensign, Friedler 2017). Without this randomness, we can propagate a human bias in our data (e.g. only patrol neighborhoods of criminals who are currently imprisoned). Unfortunately, we lack a method to understand the fairness of our predictions, other than the false positive rates of two subgroups within the population in question. One suggestion is to optimize parameter instability and disparity (Chouldechova and G'Sell 2017). More interpretably, we can perform a subset scan to detect if a given class has noteworthy bias for in a given subgroup (Zhang and Neill 2017). Such techniques only arise if we heed human biases in our data, which will be of utmost important as ML adoption continues to skyrocket.

## Societal Effects

The output of our ML systems affects real flesh and blood beings, and unfortunately, all too often, we lose sight of this reality. We focus on optimizing objectives on benchmark datasets instead of the real world applications of the code we write (Wagstaff 2012). We want to be able to transfer our expertise and models to new domains, wherein ML can augment archaic practices and automate pattern-based predictions. In example, clothing companies no longer use only intuition and actuarial science to forecast their products' performance, instead they also use ML models that incorporate seasonality, user preferences, and industry trends to decide what type of clothing should be designed in the next season (Brynjolfsson and McAffee 2014). In confluence with the proliferation of ML use cases, we must remain cognizant of the legality of our data, models, and predictions and be alert of user intent and reception.

**Current State**

Society benefits from ML models daily. These models tell us what stocks to buy, how much demand a restaurant can expect next quarter, what country poses the most threat to another, who we should date, etc. (Ross 2016). These models vicariously control us, but sometimes the mainstream media misinterprets power of ML.

In example, in the realm of natural language processing, many recent works have made bold claims that multi-agent environments have made end to end communication protocols without human intervention, which led the media to believe that AI agents make their languages and thus we need to shut them down. However, upon review, it becomes evident that language cannot emerge naturally and the systems are shut down due to a lack of interpretability: that is, one AI agent may say "Red man ball sit!" to another agent, who understands that to mean "Hello, how are you?" in English – without human intervention, the agents communicate in a nonsensical, uninterpretable grammar, basically gibberish, thus stressing the need for the first pillar (Kottur, Moura 2017).

Being aware of such misinterpretations and effectively communicating the limits of ML must be kept top of mind with mass ML adoption imminent.

**Case Study: Pricing**

In the ecommerce world, companies optimize their models to maximize their profits or increase their users' purchase frequency. One such model is a dynamic pricing engine, which prices goods based on a given consumer's willingness to pay. As such, these engines are used to serve the *optimal* price for a given user to maximize company profits. Plagued by sparse user level data and by legal constraints on what features can and cannot be used, dynamic pricing experts manage programs like time-limited coupons forecasted via a point-process model that makes real-time, global estimates based on transaction history and patterns (Manzoor and Akoglu 2017). Such pricing programs must be interpretable and unbiased; if they are not, the societal consequences of an erroneous price are catastrophic for a company. Being aware of and responsive to the implications of our models is the final key towards more humane and adoptable ML models.

## Conclusion

To be prepared for mass adoption of machine learning systems, we, as researchers and practitioners, must adopt a framework for developing humane models that ensure in-terpretability, unbiasedness, and impact. By creating a rigorous standard for machine learning interpretability, we can transform the medical predictive analytics industry. By understanding the inherent human bias in the data we collect and the sample it represents, we can ensure that we build a more unbiased model for sending prisoners back behind bars. By thinking deeply about the societal effects and ethicality of our predictions, we can ensure we deliver profitable and fair prices in the ecommerce industry. All three pillars can displace society's perception of machine learning, as the true power and beauty of how we can use autonomous agents and machine learning comes to fruition when we maintain the humanity of our models.

# Acknowledgements

Many thanks to my advisors/mentors at CMU: Jose M.F. Moura, J. Zico Kolter, and David O'Hallaron for their valuable discussions and my research partners: Satwik Kottur, Edgar Xi, and Sam Fazel for their continued support.

# References

Abah, J. 2016. Recency Bias in the Era of Big Data: The Need to Strengthen the Status of History of Mathematics In Nigerian Schools. *In Advances in Multidisciplinary and Scientific Research Journal*.

Armstrong, S., and Levinstein, B. 2017. Low Impact Artificial Intelligences. *arXiv: 1705.10720*

Bolukbasi, T., Chang, K., Zou, J., Saligrama, V., Kalai, A. 2016 Man is to Computer Programmer as Woman is to Homemaker? Debiasing Word Embeddings. *arXiv: 1607.06520*

Brynjolfsson, E., and McAffe, A. 2014. The Second Machine Age: Work, Progress, and Prosperity in a Time of Brilliant Technologies. *WW Norton & Company.*

Choi, E., Bahadori, M.T., Schuetz, A., Stewart, W.F., Sun, J. 2016. Doctor AI: Predicting Clinical Events via Recurrent Neural Networks. *In Proceedings for 2016 Machine Learning and Healthcare Conference.* Los Angeles, CA

Chouldechova, A. and G'Sell, M. 2017. Fairer and more accurate, but for whom? *In Proceedings for FAT/ML 2017.* Halifax, NS, Canada.

Doshi-Velez, F., and Kim, B. 2017. Towards A Rigorous Science of Interpretable Machine Learning. *arXiv: 1702.08608*

Doshi-Velez, F. Kortz, M, et. al. Accountability of AI Under the Law: The Role of Explanation. *arXiv: 1711.01134*

Ensign, D., Friedler, S., Neville, S., Scheidegger, C., Venkatasubramanian, S. 2017. *In Proceedings for FAT/ML 2017*. Halifax, NS, Canada.

Goodman, B. and Flaxman, S. 2016. European Union regulations on algorithmic decision-making and a "right to explanation". *In Proceedings for 2016 ICML Workshop on Human Interpretability in Machine Learning,* New York, NY.

Grace, K., Salvatier, J., Dafoe, A., Zhang, B., Evans, O. 2017. When Will AI Exceed Human Performance? Evidence from AI Experts. *arXiv: 1705.08807*

Grbovic, M., Radosavljevic, et. al. 2016. E-commerce in Your Inbox: Product Recommendations at Scale. *In Proceedings for KDD 2015*. Sydney, Australia.

Johndrow, J. and Lum, K. 2017. An algorithm for removing sensitive information: application to race-independent recidivism prediction. *arXiv: 1703.04957*

Kaplan, J. 2015. Humans Need Not Apply: A Guide to Wealth and Work in the Age of Artificial Intelligence. *Yale University Press.*

Kim, B. 2015. Interactive and interpretable machine learning models for human machine collaboration. PhD diss., Massachusetts Institute of Technology, 2015.

Kottur, S., Moura, J., Lee, S., Batra, D. 2017. Natural Language Does Not Emerge 'Naturally' in Multi-Agent Dialog. *In Proceedings for EMNLP 2017*. Denmark.

Krakovna, V. and Doshi-Velez, F. 2016. Increasing the Interpretability of Recurrent Neural Networks Using Hidden Markov Models. *In Proceedings for NIPS 2016 Workshop on Interpretable Machine Learning in Complex Systems*. Barcelona, Spain.

Lipton, Z. 2017. The Mythos of Interpretability. *In Proceedings for 2016 ICML Workshop on Human Interpretability in Machine Learning,* New York, NY.

Lipton, Z., Kale, D., Elkan, C., Wetzel, R. 2015. Learning to Diagnose with LSTM Recurrent Neural Networks. *arXiv: 1511.03677*

Manzoor, E., and Akoglu, L. 2017. RUSH! Targeted Time-limited Coupons via Purchase Forecasts. *In Proceedings for KDD 2017*. Halifax, NS, Canada.

Nigam, P. 2016. Applying Deep Learning to ICD-9 Multi-label Classification from Medical Records. *Stanford University*

O'Neil, Cathy. 2016. Weapons of Math Destruction: How Big Data Increases Inequality and Threatens Democracy. *Broadway Books.*

Ross, A. 2016. The Industries of the Future. *Simon & Schuster Paperbacks.*

Ross, A., Hughes, M., Doshi-Velez, F. 2017. Right for the Right Reasons: Training Differentiable Models by Constraining their Explanations. *arXiv:1703.03717*

Schwab, K. 2016. The Fourth Industrial Revolution. *Crown Business.*

Samek, W., Wiegand, T., Muller, K.R. 2017. Explainable Artificial Intelligence: Understanding, Visualizing and Interpreting Deep Learning Models. *arXiv: 1708.08296*

Vellido, A., Martin-Guerreo, J., Lisboa, P. 2012. Making Machine Learning Models Interpretable. *In Proceedings for European Symposium on Artificial Neural Networks, Computational Intelligence, and Machine Learning 2012*. Bruges, Belgium.

Wagstaff, K. 2012. Machine Learning that Matters. *In Proceedings for the 29th International Conference on Machine Learning.* Edinburgh, Scotland, UK

Zhao, J., Wang, T., Yatskar, M., Ordonez, V., Chang, K. 2017. Men Also Like Shopping: Reducing Gender Bias Amplification using Corpus-level Constraints. *arXiv: 1707.09457*

Zhang, Z. and Neill, D. 2017. Identifying Significant Predictive Bias in Classifiers. *arXiv: 1611.08292*